\documentclass[aps, prl, amsmath, floats, floatfix, twocolumn,superscriptaddress, nofootinbib, showpacs,reprint]{revtex4}

\usepackage{graphicx}
\usepackage{amsmath,amssymb}
\usepackage{amsfonts}
\usepackage{xspace} 
\usepackage[usenames]{color}
\usepackage{dcolumn}
\usepackage{bm}
\usepackage{mathrsfs}

\usepackage{ulem}
\definecolor {darkgreen}{rgb}{0.2,0.7,0.2}

\newcommand{\eq}{\begin{equation}}
\newcommand{\be}{\begin{equation}}
\newcommand{\eeq}{\end{equation}}
\newcommand{\ee}{\end{equation}}

\newcommand{\apjl}{Astrophys.\ J.\ Lett.\ }

\begin{document}

\title{Neutron-star mergers in scalar-tensor theories of gravity}

\author{Enrico Barausse}
\affiliation{Institut d'Astrophysique de Paris/CNRS, 98bis boulevard Arago, 75014 Paris, France}
\affiliation{Department of Physics, University of Guelph, Guelph, Ontario N1G 2W1, Canada}
\author{Carlos Palenzuela}
\affiliation{Canadian Institute for Theoretical Astrophysics, Toronto, Ontario M5S 3H8, Canada}
\author{Marcelo Ponce} 
\affiliation{Department of Physics, University of Guelph, Guelph, Ontario N1G 2W1, Canada}
\author{Luis Lehner}
\affiliation{Perimeter Institute for Theoretical Physics, Waterloo, Ontario N2L 2Y5, Canada}
\affiliation{CIFAR, Cosmology \& Gravity Program, Canada}

\begin{abstract}
Scalar-tensor theories of gravity are natural phenomenological 
alternatives to General Relativity, where the gravitational interaction is 
mediated by a scalar degree of freedom, besides the usual tensor gravitons. In regions of 
the parameter space of these theories where constraints from both solar system
experiments and binary-pulsar observations are satisfied, we show that binaries
of neutron stars present marked differences from General Relativity in both 
the late-inspiral and merger phases. In particular, phenomena
related to the spontaneous scalarization of isolated neutron stars take
place in the late stages of the evolution of binary systems, with important
effects in the ensuing dynamics. We comment 
on the relevance of our results for the upcoming Advanced LIGO/Virgo detectors. 
\end{abstract}

\date{\today \hspace{0.2truecm}}

\maketitle
General Relativity (GR) has passed stringent tests in the solar system~\cite{solar_system} 
and in binary pulsars~\cite{HT}. However, these tests involve weak gravitational fields and/or
velocities $v\ll c$, so the theory remains essentially untested in the strong-field, highly
dynamical $v\sim c$ regime, where
high-energy corrections may appear. Strong-field regimes are provided
by systems containing black holes (BHs) and/or neutron stars (NS's),
which are the target of existing (Advanced LIGO/Virgo) and future
gravitational-wave (GW) detectors. Thus the final stages of the
evolution of compact binaries provide excellent opportunities to explore gravity at extreme conditions~\cite{merger_rates}.

A natural alternative to GR is given by
scalar-tensor (ST) theories~\cite{wagoner,bergmann,nordtvedt}, where the
gravitational interaction is mediated by the usual tensor gravitons,
and by a (non-minimally coupled) scalar field. Not only are several phenomenological gravity theories
exactly equivalent to ST theories (e.g. $f(R)$ gravity~\cite{thomas_fR,wagoner}), but
a gravitational scalar (besides other degrees of freedom)
is also generally expected based on string theory.
ST theories date back to Jordan~\cite{jordan}, Fierz~\cite{fierz}, Brans and Dicke~\cite{BD}, 
and bounds have been placed on them with solar system experiments~\cite{solar_system}, 
isolated
NS's~\cite{spontaneous_scalarization,spontaneous_scalarization_bis,damour_farese_long,harada,sotani} 
and binary pulsars~\cite{eardley_dipole,ST_pulsars,will_zaglauer,BD_pulsars2}. Stricter constraints
may be obtained by detecting GWs from a gravitational collapse~\cite{collapse_ST}, from vibrating NS's~\cite{QNM_ST},
or with future-generation GW detectors~\cite{berti_buonanno_will,floating_orbits,gw_ST1,gw_ST2}. 
However, the viable parameter space of ST theories is
still sizable, so they are a rather natural choice to investigate strong-field
deviations from GR. 

We consider NS binaries and focus on strong-field/highly dynamical effects during the late inspiral/plunge 
until the merger (BHs in these theories are not expected to show
significant deviations from GR~\cite{hawking,thorne,GATech,will_latest}). 
We show that for a class of viable ST theories, NS binaries can present strong-field effects
that are qualitatively different from GR and related to the ``spontaneous scalarization'' of
isolated NS's in ST theories, first discovered in 
Refs.~\cite{spontaneous_scalarization,spontaneous_scalarization_bis} 
(see also Ref.~\cite{harada,spontaneous_scalarization_rediscovered}).
Although the merger of NS binaries is only marginally detectable with
Advanced LIGO/VIRGO in GR, for the class of viable ST theories that we consider:
(i) large deviations from GR appear which are not captured by weak-field analyses;
(ii) these effects cannot be reproduced within GR, even with an exotic equation of state;
(iii) distinct features will be detectable with Advanced LIGO/VIRGO, even in the late inspiral/plunge and merger,
unlike in GR;
(iv) these features may even have astrophysical implications in possible models for energetic electromagnetic events.

\noindent {\em Methodology:}  We consider a generic ST theory with action
\begin{equation}
\label{Jframe_action}
S=\!\!\int d^4 x\frac{\sqrt{-g}}{2\kappa}\left[\phi R-\frac{\omega(\phi)}{\phi} \partial_\mu \phi \partial^\mu \phi
\right]+ S_M[g_{\mu\nu},\psi]\,,
\end{equation}
where  $\kappa=8 \pi G$ (adopting $c=1$ throughout this Letter),
$R$ and $g$ are the Ricci scalar and determinant of the metric,
$\phi$ is the gravitational scalar, 
and $\psi$ collectively describes the matter degrees of freedom.
Although eq.~\eqref{Jframe_action} is not the most general action giving second-order field equations,
as Galileon-type terms may be present~\cite{galileon1,galileon2}, 
it includes a large family of theories
and thus provides a suitable framework for studying non-linear interactions. For instance,
Jordan-Fierz-Brans-Dicke theory corresponds to $\omega=$ const, while
$\omega(\phi)=-3/2-\kappa/(4\beta \log\phi)$  correspond 
to the theories of Refs.~\cite{spontaneous_scalarization,spontaneous_scalarization_bis}, which give
large deviations from GR for isolated NS's (``spontaneous scalarization'') and sufficiently negative $\beta$.
Also, as already mentioned, $f(R)$ gravity (both in the metric and Palatini formalism) can be remapped into
the form~\eqref{Jframe_action} (although one has to allow the presence of a potential for the scalar $\phi$)~\cite{thomas_fR,wagoner}.

One can re-express the (``Jordan-frame'') action \eqref{Jframe_action} into
the so-called ``Einstein-frame'' action through a conformal transformation $g^E_{\mu\nu}=\phi\, g_{\mu\nu}$, which yields
\begin{equation}
\label{einframe}\!
\!S=\!\!\int\!d^4 x \sqrt{-g^E} \left( \frac{R^E}{2\kappa}\!-\!\frac{1}{2}g_E^{\mu\nu} \partial_\mu\varphi \partial_\nu\varphi
\right) \! +S_M\!\!\left[\frac{g^E_{\mu\nu}}{\phi(\varphi)},\psi\right]\!\!\!
\end{equation}
where 
$\varphi$ is defined by
$({{\rm d}\log \phi}/{{\rm d}\varphi})^2={2\kappa}/[{3+2 \omega(\phi)}]$.
Imposing $\varphi=0$ for $\phi=1$, this gives
\begin{equation}
\phi=\exp\left(\sqrt{\frac{2\kappa}{3+2\omega}}\,\varphi\right)\, \, ,\quad \phi=\exp(-\beta\varphi^2)\,,
\end{equation}
respectively for Jordan-Fierz-Brans-Dicke theory and for the theories of Refs.~\cite{spontaneous_scalarization,spontaneous_scalarization_bis}.
(Our $\varphi$ is related to the scalar $\varphi_{\rm DEF}$ used
there via $\varphi=\varphi_{\rm DEF}/\sqrt{4\pi G}$.)

In the Einstein frame the field equations are
\begin{gather}\label{einstein}
G_{\mu\nu}^{E}=\kappa \left(T^\varphi_{\mu\nu} + T^E_{\mu\nu} \right),\\
\Box^E \varphi = \frac12 
\frac{{\rm d}\log \phi}{{\rm d}\varphi} T_E\label{KG} \,,\\
\nabla_\mu^E T_E^{\mu\nu}=-\frac{1}{2} T_E 
\frac{{\rm d}\log \phi}{{\rm d}\varphi} 
g_E^{\mu\nu} \partial_\mu \varphi\,,\label{scalarT}
\end{gather}
where 
\begin{eqnarray} 
T_E^{\mu\nu}&=&\frac{2}{\sqrt{-g^E}}\frac{\delta S_M}{\delta g^E_{\mu\nu}} \,\,\,\,\, \mbox{and} \\
T^\varphi_{\mu\nu} &=&\partial_\mu \varphi \partial_\nu \varphi- \frac{g^E_{\mu\nu}}{2} g_E^{\alpha\beta} \partial_\alpha \varphi
\partial_\beta \varphi\, 
\end{eqnarray} 
are the matter and scalar-field stress-energy tensors in the Einstein frame, and $T_E\equiv T_E^{\mu\nu}g^E_{\mu\nu}$.
Indices 
are raised/lowered with the Einstein-frame metric $g_E$, and the relation 
between the stress-energy tensors in the two frames is
$
T_E^{\mu\nu}={T^{\mu\nu}}{\phi^{-3}} \, \, , \, \, T^E_{\mu\nu}={T_{\mu\nu}}{\phi^{-1}}.
$
Also,
$u^\mu=\sqrt{\phi}\,u_E^\mu $ (from $g^E_{\mu\nu} u_E^\mu u_E^\nu=-1$); 
$\rho=\phi^2 {\rho}_E$ (from $\rho_E= u_E^\mu u_E^\nu T^E_{\mu\nu}$) and  $p=\phi^2 {p}_E$
(from $T=\phi^2 T_E$). Last, to preserve the same equation of state in both frames, the rest-mass densities must be related by $\rho_0=\phi^2 {\rho}^E_0$. 
Baryon number conservation in the Jordan frame ($\nabla_\mu j^\mu=0$ with $j^\mu=\rho_0 u^\mu$) then
gives
\eq\label{rest_mass_cons_Einstein}
\nabla^E_\mu j_E^\mu = 
-\frac{1}{2} 
\frac{{\rm d}\log \phi}{{\rm d}\varphi}
j_E^\mu \partial_\mu \varphi \,,
\eeq
with $j_E^\mu={\rho}^E_0 u_E^\mu$.
Therefore, solving the system~\eqref{einstein},~\eqref{KG},~\eqref{scalarT} and~\eqref{rest_mass_cons_Einstein}
and transforming back to the Jordan frame provides a solution to the original problem. We adopt 
this approach here.

\noindent {\em Physical Set-up:}
We model the NS's with a perfect fluid coupled to the full field
equations~(\ref{einstein}-\ref{scalarT},
\ref{rest_mass_cons_Einstein}) to accurately represent the strong
gravitational effects during the binary's evolution.
Our techniques for solving these equations have been
 described and
tested previously~\cite{Anderson:2006ay,Palenzuela:2006wp,Liebling,Anderson:2007kz,Anderson:2008zp,had_webpage}.
The initial data are evolved in a cubical computational domain 
$x^i \in [-350,350]$ km, and we employ an adaptive mesh refinement
that tracks the compact objects with cubes slightly larger
than their radii and resolution $\Delta x = 0.5$ km.
We consider an unequal-mass binary system, initially on a quasi-circular
orbit with separation of $60$ km and angular velocity 
$\Omega=1295$ rad/s, constructed with {\sc Lorene}~\cite{lorene_webpage}.
The stars are described by a polytropic equation of state ($p/c^2=K \rho_0^\Gamma$) with $\Gamma=2$ and $K=123 G^3 M_\odot^2/c^6$
(which yields a maximum ADM mass of about $1.8 M_\odot$ both in GR and in the ST theories we consider).
We adopt a mass ratio of $q \equiv 0.937$, possible for progenitors of gamma-ray bursts~\cite{GRB_mass_ratio},
and choose individual baryon masses $\{1.78, 1.90\} M_\odot$, corresponding to 
gravitational masses $\{ 1.58, 1.67\} M_\odot$. 

\begin{figure}
\includegraphics[width=4.9cm,,angle=-90]{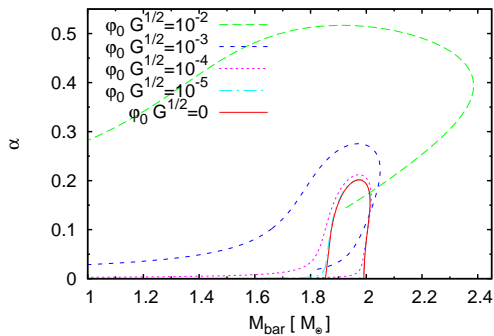}
\vskip-0.3cm
\caption{\footnotesize Effect of an external scalar field $\varphi_0$, for $\beta/(4\pi G)=-4.5$.\label{fig:scalarization}}
\end{figure}

For the gravity theory, we consider 
$\omega(\phi)=-3/2-\kappa/(4\beta \log\phi)$ [corresponding to $\phi=\exp(-\beta\varphi^2)$].
As mentioned, these theories are equivalent to those of Refs.~\cite{spontaneous_scalarization,spontaneous_scalarization_bis}.
Besides the constant $\beta$, the gravity theory is also characterized 
by the asymptotic value $\varphi_0$ of the scalar~\cite{spontaneous_scalarization}. 
Binary pulsar measurements require $\beta/(4\pi G)\gtrsim -4.5$~\cite{ST_pulsars}, while the Cassini experiment 
constrains $\varphi_0<\varphi_{_{\rm Cassini}}\equiv
2(G\pi)^{1/2}/[|\beta| (3+2 \omega_0)^{1/2}]\approx1.26\times10^{-2}G^{1/2}/|\beta|$
 (with $\omega_{0}=4\times10^4$~\cite{ST_pulsars,solar_system}). 
Moreover,
from $\beta/(4\pi G)\sim -4$ to $\beta/(4\pi G)= -4.5$, the allowed value for $\varphi_0$ decreases
 from $\varphi_{_{\rm Cassini}}$ to $0$, again due to
constraints from binary pulsars~\cite{ST_pulsars}.  
For the system described above, we tried various values of  $\varphi_0\leq 10^{-5} G^{-1/2}$, and the results
do not change significantly when $\beta$ is fixed. (As will become clear from Fig.~\ref{fig:scalarization} and
associated discussion, larger
values of $\varphi_0$, even when allowed by existing constraints, induce even larger deviations from GR.)

\noindent {\em GW extraction and backreaction:} The response of a GW detector is encoded in the
curvature scalars in the physical (Jordan) frame~\cite{eardley_dof}. These are obtained from the Einstein 
frame components as $\psi_4 =- R_{\ell \bar{m}\ell \bar{m}}=\phi\, \psi_4^E$, $\psi_3 =-R_{\ell k\ell \bar{m}}/2=\phi \,\psi_3^E+ ...$,
$\psi_2=- R_{\ell k\ell k}/6=\phi \,\psi_2^E+...$ and
$\phi_{22}=- R_{\ell {m}\ell \bar{m}}=\phi\, \left(\phi_{22}^E- l^\nu l^\mu \nabla_{\nu} \nabla_{\mu} \log \phi/2 + ...\right)$
(with $...$ denoting subleading terms in the distance to the detector and $l,m$ being components of a null tetrad adapted to outgoing wavefronts).
Far from the source one expects $\varphi=\varphi_0+\varphi_1/r+...$, with $\varphi_0=$ const 
and $\varphi_1$ a function of $x^\mu$, so because of the peeling property in the Einstein frame, $\psi_2$ and $\psi_3$ decay faster than $1/r$
and do not produce observable effects on a GW detector at infinity. 
However, using 
$\log \phi = -\beta \varphi^2=- \beta (\varphi_0^2+2 \varphi_0 \varphi_1/r)+...$,
one 
obtains $\phi_{22}\sim \beta \varphi_0 \partial_t^2\varphi_1/r$. Thus, 
the radiative degrees of freedom (decaying as $1/r$ and observable by GW detectors) are 
$\psi_4$ (i.e. tensor gravitons) and $\phi_{22}\sim \beta \varphi_0\partial_t^2\varphi_1/r$ (i.e. a purely transverse, 
radiative scalar 
mode~\cite{eardley_dof}).
Nonetheless, for  $\varphi_0\to0$ the $1/r$ radiative component of $\phi_{22}$ vanishes. 
Thus, since $\varphi_0$ is constrained to small values, for viable ST theories the scalar 
mode couples weakly to 
GW detectors~\cite{ST_pulsars}, which   
makes its direct detection problematic.

In fact, it is easy to get convinced, already at the level of the
action \eqref{einframe}, that the scalar mode is not observable directly in the limit $\varphi_0\to0$.
The detection of GWs is based
 on free-falling test masses, so to analyze the detector's response one needs to look at the Jordan frame metric $g^E_{\mu\nu}/\phi(\varphi)$,
to which the matter fields $\psi$ couple [cf. eq. \eqref{einframe}]. Far from the source, in suitable coordinates one has
$g^E_{\mu\nu}\approx\eta_{\mu\nu}+h_{\mu\nu}$ and $\varphi\approx\varphi_0+\delta\varphi$, where $h_{\mu\nu}$
and $\delta\varphi$ are small perturbations. If $\varphi_0=0$, we have $\phi=\exp(-\beta\varphi^2)\approx1-\beta\delta\varphi^2$,
and therefore $g^E_{\mu\nu}/\phi(\varphi)\approx \eta_{\mu\nu}+h_{\mu\nu}$ at linear order. This means that the motion
of the detector's
test masses is only sensitive to the tensor waves $h_{\mu\nu}$ in the limit $\varphi_0\to0$.

\begin{figure}
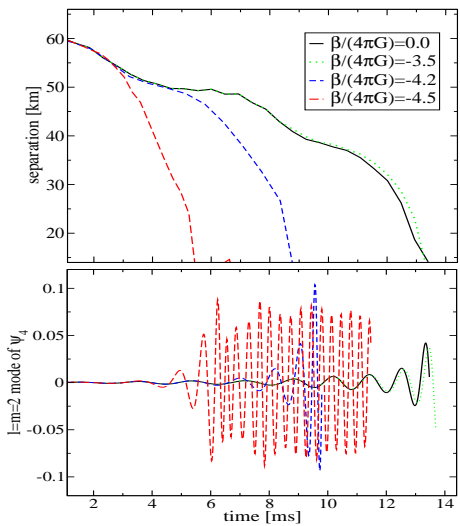

\centering
\includegraphics[height=3.4cm,width=5.77cm,angle=0]{separation.eps}\\
\vskip 0.028cm
\hskip -0.25 cm\includegraphics[height=3.4cm,width=6.cm,angle=0]{c22.eps}
\caption{\footnotesize The separation and the dominant mode of the $\psi_4$ scalar (encoding the effect of GWs)
for a binary with gravitational masses $\{ 1.58, 1.67\} M_\odot$, and for different values of $\beta$.
\label{fig:separation_gw}}
\end{figure}

Still, although weakly coupled to a GW detector at infinity, the scalar mode carries energy away from the source
[cf. eqs.~\eqref{KG} and~\eqref{scalarT}]
and exerts a significant backreaction on it,
because the scalar fluxes appear at 1.5PN order,
while the quadrupolar tensor fluxes of GR appear at 2.5 PN.
More precisely, for a quasicircular binary with masses $m_1$ and $m_2$, and scalar charges $\alpha_1$ and $\alpha_2$ [with 
$\alpha_i\approx\sqrt{4\pi/G}\,\varphi^i_1/m_i$,
where $\varphi_1$ is defined, as above, by $\varphi=\varphi_0+\varphi_1/r+...$],
the dipolar scalar emission is~\cite{eardley_dipole,will_zaglauer,damour_farese_long}
\begin{equation}\label{dipole_flux}
\dot{E}_{\rm dipole} = \frac{G}{3 c^3} \left(\frac{G_{\rm eff} m_1 m_2}{r^2}\right)^2 (\alpha_1-\alpha_2)^2\,.
\end{equation}
Here, $G_{\rm eff}=G (1+\alpha_1 \alpha_2)$ is
the effective gravitational constant appearing in the Newtonian interaction between the stars, 
i.e. the gravitational
force gets modified by the exchange of scalar gravitons and becomes~\cite{damour_farese_long}
\begin{equation}\label{modifiedF}
F= \frac{G_{\rm eff} m_1 m_2}{r^2}\,.
\end{equation}
The quadrupole tensor emission is instead~\cite{will_zaglauer,damour_farese_long}
 \begin{equation}
\dot{E}_{\rm quadrupole} = \frac{32 G}{5 c^3} \left(\frac{G_{\rm eff} m_1 m_2}{r^2}\right)^2
                           \left( \frac{v}{c} \right)^2\,,
\end{equation}
where $v=[G_{\rm eff} (m_1+m_2)/r]^{1/2}$ is the relative velocity of the two stars.
Therefore, the dipolar scalar fluxes are produced abundantly during the inspiral 
if the charges $\alpha_1$ and $\alpha_2$ are different,
and dominate over the tensor quadrupole fluxes, which are suppressed by $(v/c)^2$ relative to them.

\noindent {\em Results and comparison to GR:}
Our simulations confirm the qualitative features described above, but also highlight 
a more intricate phenomenology. 
Specifically, in ST theories 
with $\beta/(4\pi G)\lesssim - 4.2$, NS
binaries merge at significantly lower frequency than in GR, e.g. in  Fig.~\ref{fig:separation_gw}
the plunge starts 
already when the stars' centers are $\sim52$ km apart,
corresponding to an angular velocity 
$\Omega \sim 1850$ rad/s (i.e. 
a GW frequency $f \sim 586$ Hz, within  Advanced LIGO/Virgo's sensitivity bands),
and results in the formation of a rotating bar (whose long-lived GW signal is seen in the lower panel).  
Remarkably, plunges starting so early cannot be
obtained in GR, because even with exotic equations of state, NS radii
are constrained to $R_{\rm NS}\lesssim 14$ km~\cite{eos}, so the interaction between the
two stars does not trigger a plunge until a separation $\sim 2R_{\rm NS}\lesssim 28$ km.
Clearly, because a NS binary spends a large part of its inspiral within
LIGO/Virgo's sensitivity bands, these early plunges will not produce a signal-to-noise ratio
very different from GR and will not jeopardize the source's detection. Given the magnitude of 
the differences highlighted in Fig.~\ref{fig:separation_gw} and the fact that they appear well within
advanced detectors' frequency windows, however, it appears likely
that a suitable post-detection analysis (i.e. at the parameter-estimation stage) will be able to highlight them. 
(A more detailed analysis of this point goes beyond the scope of this paper, and will be presented
elsewhere.)

\begin{figure}
\centering
\includegraphics[height=3.6cm,width=4.cm,angle=0]{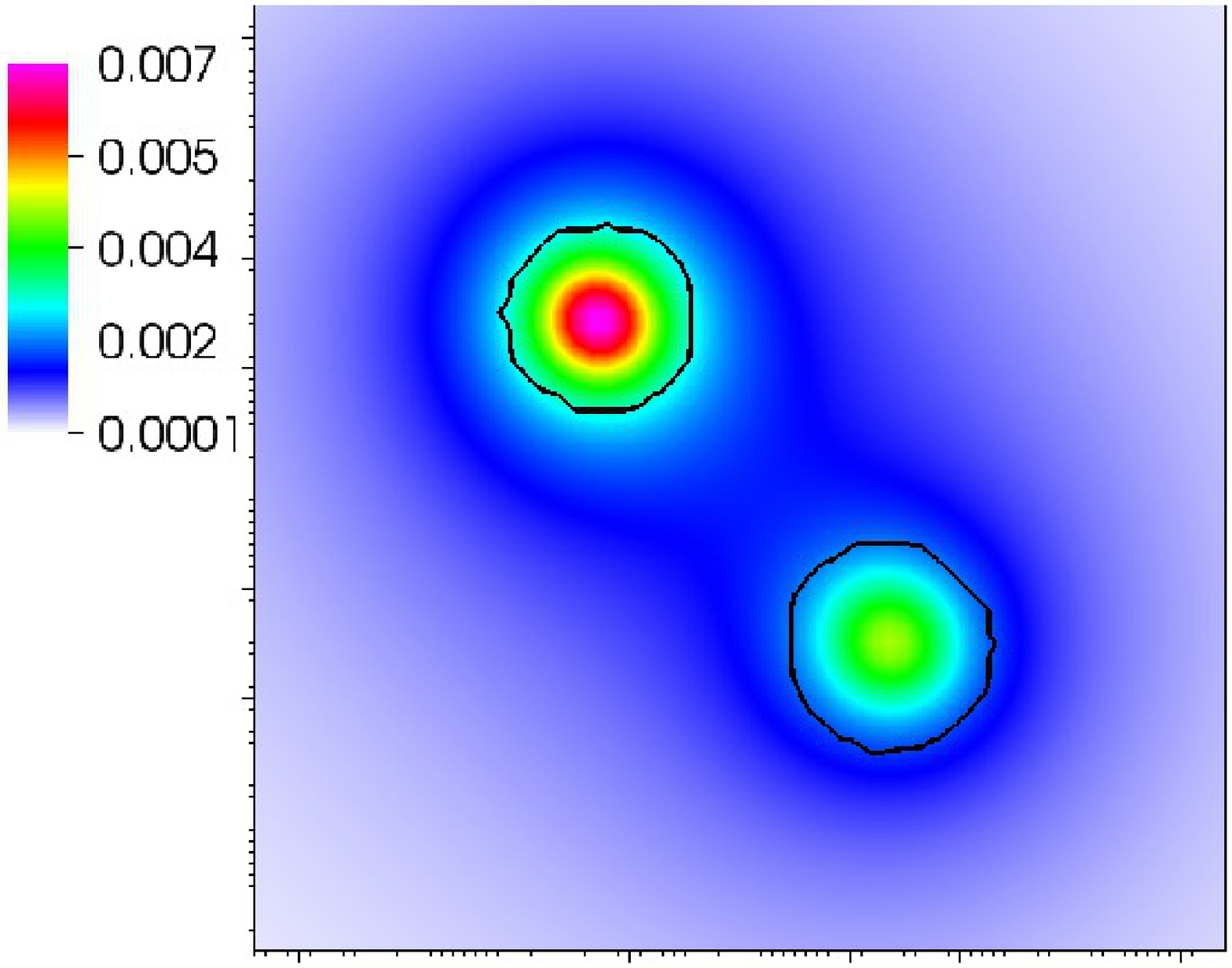}
\includegraphics[height=3.6cm,width=4.cm,angle=0]{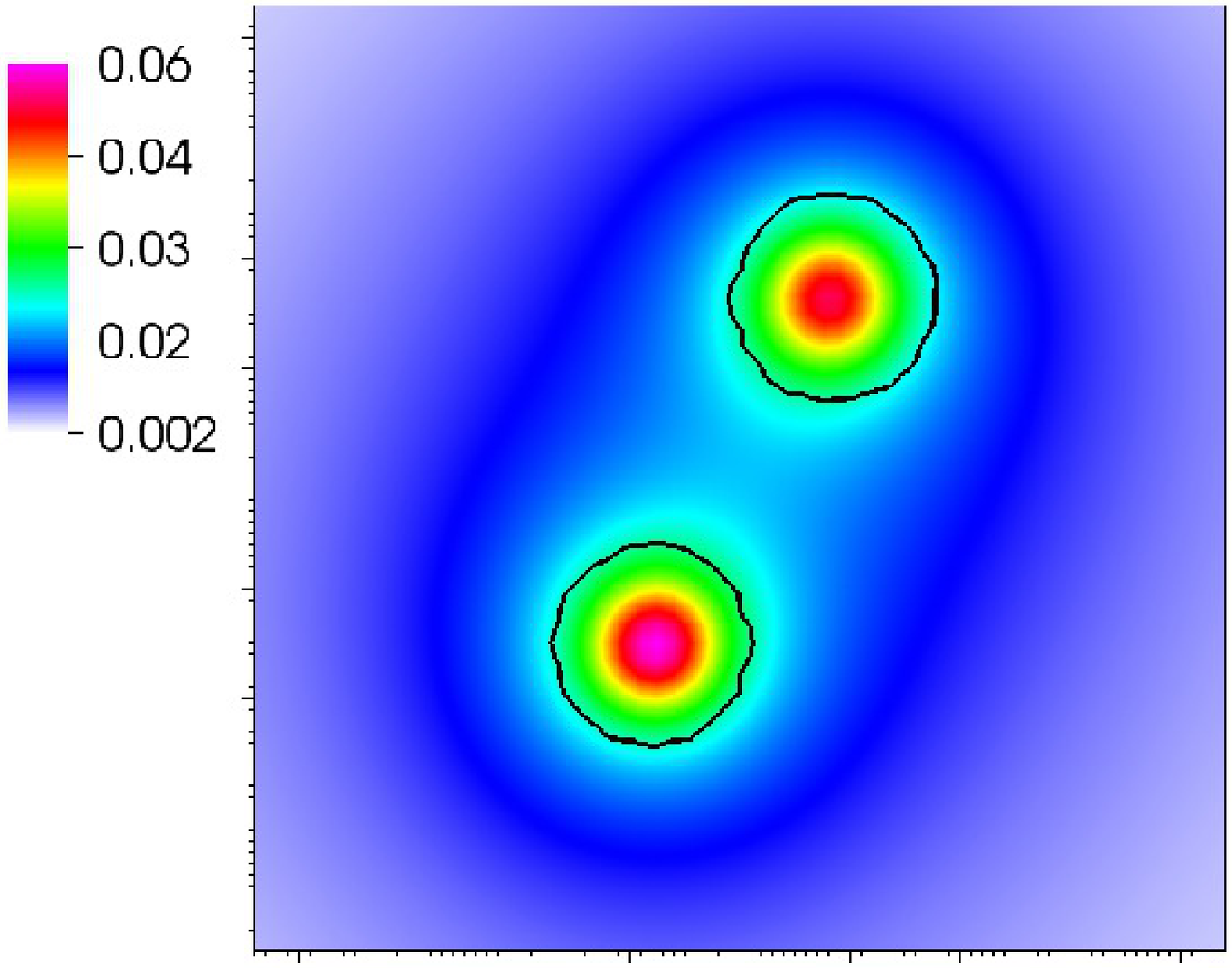}\\
\vskip -0.6cm
\includegraphics[height=3.6cm,width=4.cm,angle=0]{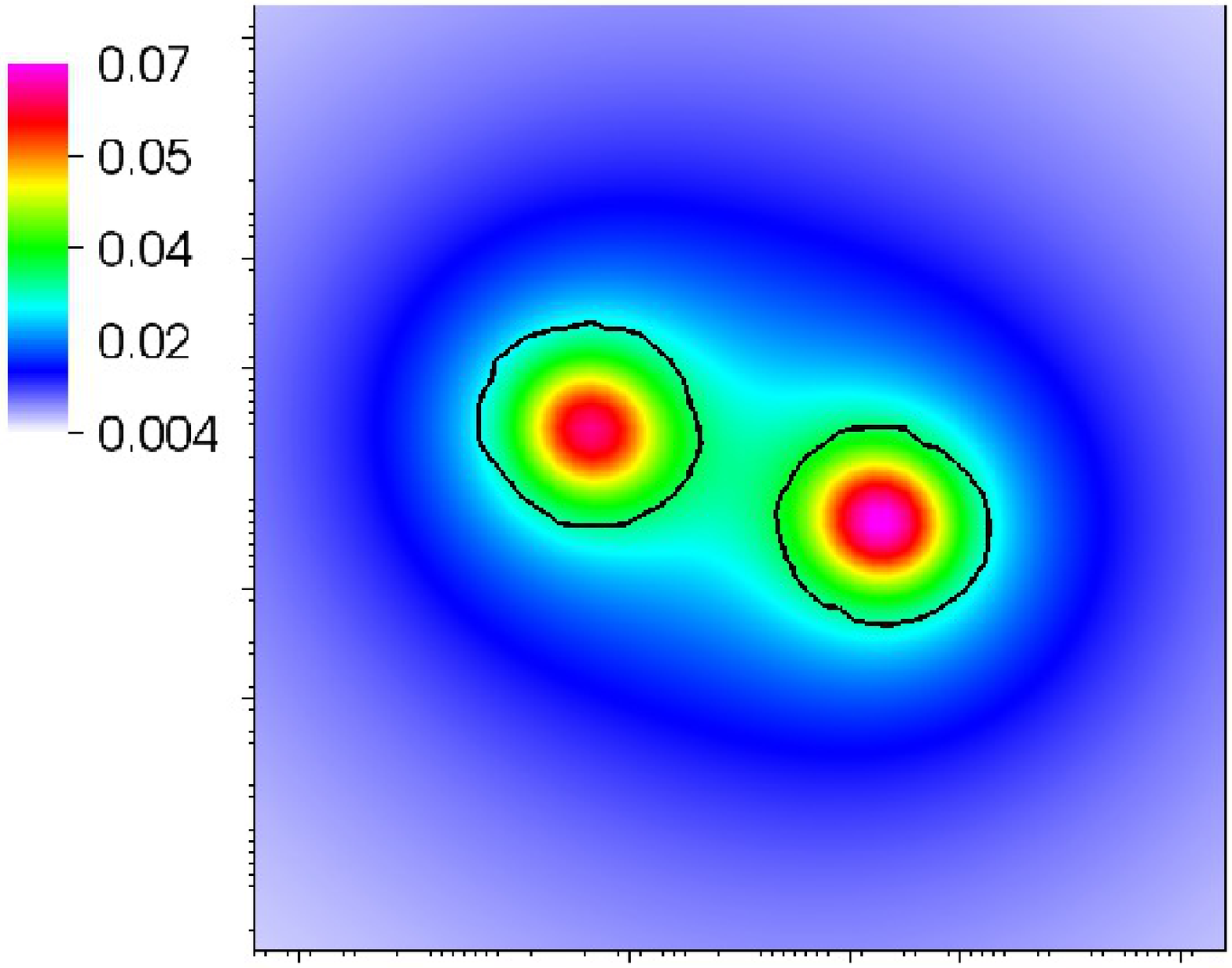}
\includegraphics[height=3.6cm,width=4.cm,angle=0]{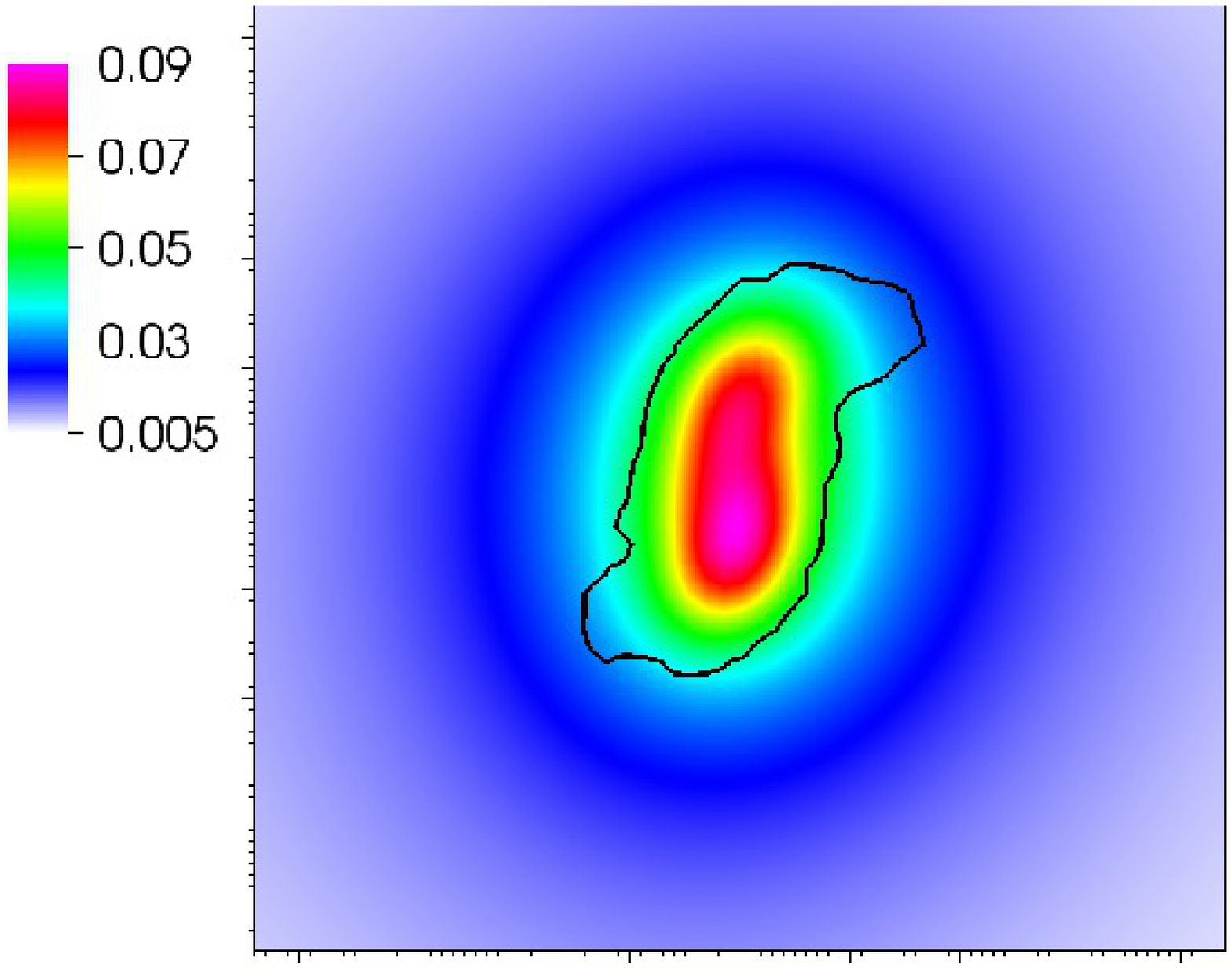}
\vskip -0.3cm
\caption{\footnotesize The scalar field $\varphi G^{1/2}$ (color code) and the NS surfaces
(solid black line) at $t=\{1.8,3.1,4.0,5.3\}$ms 
for $\beta/(4 \pi G) = -4.5$, and the binary of Fig.~\ref{fig:separation_gw}.
\label{snapshots}}
\end{figure}

The cause of these earlier mergers is not simply the backreaction of the scalar
 fluxes~\eqref{dipole_flux} (absent in GR).
In fact, even
though our initial data essentially 
maximize the dipolar emission~\eqref{dipole_flux}
by giving the first star a charge close to the maximum value allowed by the ST theory
($\alpha_1\sim\alpha_{\rm max}$), and
an almost zero scalar charge to the second star ($\alpha_2\approx0$), 
the scalar field grows rapidly inside the
second star, which quickly develops a charge $\alpha_2\approx\alpha_1$ when the binary becomes
 sufficiently close (cf. Fig.~\ref{snapshots}). This shuts off the dipolar flux~\eqref{dipole_flux},
but enhances the Newtonian pull~\eqref{modifiedF}. Therefore,
the earlier mergers are caused by
the combination of dissipative [eq.~\eqref{dipole_flux}] and conservative [eq.~\eqref{modifiedF}] effects.
 As a qualitative test, we integrated the PN equations of
motion of GR with $G$ replaced by $G_{\rm eff}=G (1+\alpha_1 \alpha_2)$ 
[so as to mimic eq.~\eqref{modifiedF}, with $\alpha_1,\,\alpha_2\sim 0.2-0.4$
set to values compatible with our simulations], and confirmed that the enhanced 
gravitational pull induces quicker mergers.

The growth of the scalar field and charge of 
non-scalarized stars getting close
to scalarized ones can be understood in simple terms. 
The scalar field extends beyond the radius of the baryonic matter~\cite{spontaneous_scalarization,spontaneous_scalarization_bis}. 
Indeed, defining an effective radius $L$ for the scalar as that enclosing a fixed fraction, e.g. 90\%,
of its mass, one gets $L/R_{\rm NS} \sim 4-5$ for isolated stars (cf. also Fig.~\ref{snapshots}).
When the non-scalarized star
enters this scalar ``halo'' of the scalarized star, it grows a significant
 charge. This can be seen
by studying isolated NS's~\cite{spontaneous_scalarization,spontaneous_scalarization_bis}, and imposing a
 non-zero asymptotic value $\varphi_0$ for the scalar 
field, in order to mimic
the effect of the ``external'' scalar field produced by the other (scalarized) star. The 
effect of $\varphi_0$ is shown in
Fig.~\ref{fig:scalarization}, where we used a static, spherically symmetric code to calculate the scalar 
charge of NS's as a function of the baryonic mass, for a ST theory with $\beta/(4\pi G)=-4.5$. As
can be seen, even modest values of $\varphi_0$
induce significant scalar charges.
This phenomenon, known
as ``induced scalarization''~\cite{spontaneous_scalarization,spontaneous_scalarization_bis,damour_farese_long},
has also been observed for boson stars in ST theory~\cite{induced_scalarization_boson_stars}, and
is similar, energetically,
to the magnetization of a ferromagnetic material 
in a sufficiently strong magnetic field~\cite{spontaneous_scalarization,spontaneous_scalarization_bis,ferromagnets}. Here, 
the external 
scalar field makes the configuration with non-zero charge energetically preferred over the initial non-charged one.

\begin{figure}
\centering
\includegraphics[height=3.7cm,width=6.2cm,angle=0]{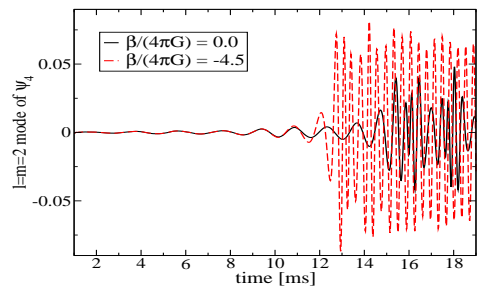}
\caption{\footnotesize The dominant mode of the $\psi_4$ scalar,
with $\beta/(4\pi G)=-4.5$, $\varphi_0=10^{-5} G^{-1/2}$,
for an equal-mass binary with individual baryon masses of $1.625 M_\odot$.
\label{fig:gw2}}
\end{figure}

Quite remarkably, the growth of the scalar field inside stars that are sufficiently close seems quite robust,
(though its magnitude naturally depends on the values of $\beta$ and $\varphi_0$).
In fact, it happens also in systems where induced scalarization is likely unable to trigger the scalar's initial growth, e.g.
in (at least) some binaries whose stars are initially non-scalarized,
and far from the ``critical mass'' $M_{\rm bar}\approx 1.85 M_\odot$ marking the onset
of spontaneous scalarization for small $\varphi_0$ in Fig.~\ref{fig:scalarization}.
For instance, in Fig.~\ref{fig:gw2} we show the waveforms for an equal-mass binary whose
stars have baryon mass 1.625 $M_\odot$, gravitational mass $1.47 M_\odot$
and radius $R_{\rm NS}=13$ km,
for GR and a ST theory with $\beta/(4\pi G)=-4.5$ and $\varphi_0 \sqrt{G}=10^{-5}$. Clear deviations from GR
arise at $t \sim 10$ ms, 
corresponding to a separation $R\sim 40$ km and $f\sim 645$ Hz.
These deviations will occur at later (earlier) times for smaller (larger) $\varphi_0$.
We will study these
smaller-mass systems more in future work, but 
this result is not entirely surprising. The spontaneous scalarization
of isolated stars occurs when a non-zero value $\varphi_c$ of the scalar at the center becomes energetically
favored over $\varphi_c=0$. Refs.~\cite{spontaneous_scalarization,spontaneous_scalarization_bis,ferromagnets} noted indeed that the 
star's energy is $E\sim \int [(\nabla \varphi)^2/2+\rho \exp(\beta \varphi^2)] d^3 x \sim
\varphi_c^2 L+ M \exp(\beta \varphi_c^2)$, where the length $L\sim |\varphi/\nabla\varphi|$
regulates the scalar's gradients. [As mentioned, $L\sim 4-5 R_{\rm NS}$, because $\varphi$
decays slowly ($\sim1/r$) outside the star]. 
One can easily check that if $M/L$ is large enough (i.e., if the star is compact enough) and $\beta<0$, 
$E$ may have a  minimum at $\varphi_c\neq0$, and the star will spontaneously scalarize. 
In a tight binary, however, the scalar
field will change on the scale of the  
separation $R$ (cf. Fig.~\ref{snapshots}), so $E\sim \varphi_c^2 R+ M \exp(\beta \varphi_c^2)$,
suggesting that at sufficiently small separations, the
energy's minimum will lie at $\varphi_c\neq0$, i.e. $\varphi$ may grow
inside stars that would not scalarize spontaneously in isolation.

Finally, our findings might have implications for short gamma-ray bursts,
of which NS binaries are likely progenitors.
While restricted to ST theories with $\beta/(4\pi G)\lesssim -4.2$
(and possibly to binary masses $M_{\rm ADM}\gtrsim 3 M_\odot$), 
our results show that in principle, modifications to the gravity
theory may cause the GW signal and the orbital evolution to differ from GR.
This may be important for coincident
searches of GW and electromagnetic signals from NS binaries, and 
 for energetic events possibly associated with NS mergers and
their after-merger remnant (because the extra scalar
channel carrying energy away from the binary can affect
its energy budget).

\textit{Acknowledgments:} We are deeply indebted to Gilles Esposito-Farese for providing insightful 
suggestions on different physical effects, observables and on the spontaneous 
scalarization induced on a star by another in ST theory. Also, we thank our collaborators
M. Anderson, E. Hirschmann, S.L. Liebling  and D. Neilsen with whom we have developed the basic computational
infrastructure employed in this work. 
We acknowledge support from a CITA National Fellowship while
at the University of Guelph, and from
the European Union's Seventh Framework Programme (FP7/PEOPLE-2011-CIG)
through the  Marie Curie Career Integration Grant GALFORMBHS PCIG11-GA-2012-321608
while at the Institut d'Astrophysique de Paris (to E.B.); the Jeffrey L.~Bishop Fellowship (to C.P.)
and NSERC through a Discovery grant (to L.L.). We also acknowledge hospitality
from the Kavli Institute for Theoretical Physics (UCSB), where part of this work was
carried out. This work was supported in part by the
National Science Foundation under grant No. NSF PHY11-25915.
Computations were performed on Scinet.
Research at Perimeter Institute is supported through Industry Canada
and by the Province of Ontario through the Ministry of Research and Innovation.


\begin{thebibliography}{90}

\bibitem{solar_system} 
  C.~M.~Will,
  Living Rev.\ Rel.\  {\bf 4}, 4 (2001)
  [gr-qc/0103036].
\bibitem{HT}  J.~H.~Taylor and J.~M.~Weisberg, J.~M. \apj {\bf 253}, 908 (1982); T.~Damour and J.~H.~Taylor,
  Phys.\ Rev.\ D {\bf 45}, 1840 (1992).

 \bibitem{merger_rates}   
  J.~Abadie {\it et al.}  [LIGO Scientific and Virgo Collaborations],
  Class.\ Quant.\ Grav.\  {\bf 27}, 173001 (2010).
 
\bibitem{wagoner} R.~Wagoner, Phys.\ Rev.\ {\bf D} 1, 3209 (1970).
\bibitem{bergmann} P.~G.~Bergmann, Int.\ J.\ Theor.\ Phys.\ {\bf 1}, 25 (1968).
\bibitem{nordtvedt} K.~Nordtvedt, \apj {\bf 161}, 1059 (1970)
\bibitem{thomas_fR}   T.~P.~Sotiriou,
  Class.\ Quant.\ Grav.\  {\bf 23}, 5117 (2006).
\bibitem{jordan} P.~Jordan, Z. Phys. {\bf 157}, 112 (1959)
\bibitem{fierz} M.~Fierz, Helv. Phys. Acta {\bf 29}, 128 (1956)
\bibitem{BD} C.~Brans and R.~H.~Dicke, Phys.\ Rev.\ {\bf 124}, 925 (1961)

%

  
  
  
  

  
  
    
  
  
%

%
%
%
%
%
  

%
\bibitem{spontaneous_scalarization} T.~Damour and G.~Esposito-Farese, Phys.\ Rev.\ Lett. {\bf 70}, 2220 (1993)
 \bibitem{spontaneous_scalarization_bis}  T.~Damour and G.~Esposito-Farese,
  Phys.\ Rev.\ D {\bf 54}, 1474 (1996)
 \bibitem{damour_farese_long}  T.~Damour and G.~Esposito-Farese,  Class.\ Quant.\ Grav.\  {\bf 9}, 2093 (1992).
 \bibitem{harada}   T.~Harada,
  Phys.\ Rev.\ D {\bf 57}, 4802 (1998)
 \bibitem{sotani}  H.~Sotani,
  Phys.\ Rev.\ D {\bf 86}, 124036 (2012)




\bibitem{eardley_dipole} D.~M.~Eardley, \apjl {\bf 196}, L59 (1975)

\bibitem{ST_pulsars}  T.~Damour and G.~Esposito-Farese, Phys.\ Rev.\ D {\bf 58}, 042001 (1998);  
P.~C.~C.~Freire, N.~Wex, G.~Esposito-Farese, J.~P.~W.~Verbiest, M.~Bailes, B.~A.~Jacoby, M.~Kramer and I.~H.~Stairs {\it et al.},
  Mon.\ Not.\ Roy.\ Astron.\ Soc.\  {\bf 423}, 3328 (2012)
\bibitem{will_zaglauer}   C.~M.~Will and H.~W.~Zaglauer,
  Astrophys.\ J.\  {\bf 346}, 366 (1989)


\bibitem{BD_pulsars2}  
  J.~Alsing, E.~Berti, C.~M.~Will and H.~Zaglauer,
  Phys.\ Rev.\ D {\bf 85}, 064041 (2012)
  
\bibitem{collapse_ST}   J.~Novak,  Phys.\ Rev.\ D {\bf 57}, 4789 (1998).
\bibitem{QNM_ST}   H.~Sotani and K.~D.~Kokkotas,
  Phys.\ Rev.\ D {\bf 70}, 084026 (2004)
  
\bibitem{berti_buonanno_will}  C.~M.~Will and N.~Yunes,
  Class.\ Quant.\ Grav.\  {\bf 21}, 4367 (2004);
E.~Berti, A.~Buonanno and C.~M.~Will,
  Phys.\ Rev.\ D {\bf 71} (2005) 084025  

\bibitem{floating_orbits}    V.~Cardoso, S.~Chakrabarti, P.~Pani, E.~Berti and L.~Gualtieri,
  Phys.\ Rev.\ Lett.\  {\bf 107}, 241101 (2011);   N.~Yunes, P.~Pani and V.~Cardoso,  Phys.\ Rev.\ D {\bf 85}, 102003 (2012).  
\bibitem{gw_ST1}    E.~Berti, L.~Gualtieri, M.~Horbatsch and J.~Alsing,
  Phys.\ Rev.\ D {\bf 85}, 122005 (2012).

\bibitem{gw_ST2}  
  K.~G.~Arun,
  Class.\ Quant.\ Grav.\  {\bf 29}, 075011 (2012).

\bibitem{hawking} S.~W.~Hawking, Comm.\ Math.\ Phys.\ {\bf 25}, 167 (1972) 
\bibitem{thorne} K.~S.~Thorne and J.~J. Dykla, \apjl {\bf 166}, L35 (1971) 
\bibitem{GATech} 
  J.~Healy, T.~Bode, R.~Haas, E.~Pazos, P.~Laguna, D.~M.~Shoemaker and N.~Yunes,
  arXiv:1112.3928 [gr-qc].
\bibitem{will_latest}   S.~Mirshekari and C.~M.~Will,
  arXiv:1301.4680 [gr-qc].


 \bibitem{spontaneous_scalarization_rediscovered}  P.~Pani, V.~Cardoso, E.~Berti, J.~Read and M.~Salgado,
  Phys.\ Rev.\ D {\bf 83}, 081501 (2011)

\bibitem{galileon1}  G.~W.~Horndeski, Int.\ J.\ Theor.\ Phys.\  {\bf 10}, 363 (1974).
\bibitem{galileon2} C.~Deffayet, X.~Gao, D.~A.~Steer and G.~Zahariade,
  Phys.\ Rev.\ D {\bf 84}, 064039 (2011)



\bibitem{Anderson:2006ay} 
  M.~Anderson, E.~Hirschmann, S.~L.~Liebling and D.~Neilsen,
  Class.\ Quant.\ Grav.\  {\bf 23}, 6503 (2006)
  [gr-qc/0605102].

\bibitem{Palenzuela:2006wp} 
  C.~Palenzuela, I.~Olabarrieta, L.~Lehner and S.~L.~Liebling,
  Phys.\ Rev.\ D {\bf 75}, 064005 (2007).

\bibitem{Liebling} 
  S.~L.~Liebling,
  Phys.\ Rev.\ D {\bf 66}, 041703 (2002).

\bibitem{Anderson:2007kz} 
  M.~Anderson, E.~W.~Hirschmann, L.~Lehner, S.~L.~Liebling, P.~M.~Motl, D.~Neilsen, C.~Palenzuela and J.~E.~Tohline,
  Phys.\ Rev.\ D {\bf 77}, 024006 (2008).

\bibitem{Anderson:2008zp} 
  M.~Anderson, E.~W.~Hirschmann, L.~Lehner, S.~L.~Liebling, P.~M.~Motl, D.~Neilsen, C.~Palenzuela and J.~E.~Tohline,
  Phys.\ Rev.\ Lett.\  {\bf 100}, 191101 (2008)

\bibitem{had_webpage}
\bibinfo{note}{{http://www.had.liu.edu/}}.


\bibitem{lorene_webpage}
\bibinfo{note}{{http://www.lorene.obspm.fr}}.



  

\bibitem{GRB_mass_ratio} 
  K.~Belczynski, R.~W.~O'Shaughnessy, V.~Kalogera, F.~Rasio, R.~Taam and T.~Bulik,
  \apjl {\bf 680} L129  (2008)

  
\bibitem{eardley_dof} D.~M.~Eardley, D.~L.~Lee and A.~P.~Lightman, Phys.\ Rev.\ D  {\bf 8}, 3308 (1973)

\bibitem{eos}  A.~W.~Steiner, J.~M.~Lattimer and E.~F.~Brown,
  Astrophys.\ J.\  {\bf 722}, 33 (2010)

\bibitem{induced_scalarization_boson_stars}  M.~Ruiz, J.~C.~Degollado, M.~Alcubierre, D.~Nunez and M.~Salgado,
  Phys.\ Rev.\ D {\bf 86}, 104044 (2012)


\bibitem{ferromagnets} G.~Esposito-Farese,  AIP Conf.\ Proc.\  {\bf 736}, 35 (2004)
  [gr-qc/0409081].


\end{thebibliography}
\end{document}